# Active-Learning in the Online Environment

Zahra Derakhshandeh
California State University, East Bay
zahra.derakhshandeh@csueastbay.edu

Babak Esmaeili
Arizona State University
besmaeil@asu.edu

## ABSTRACT

Online learning is convenient for many learners; it gives them the possibility of learning without being restricted by attending a particular classroom at a specific time. While this exciting opportunity can let its users manage their life in a better way, many students may suffer from feeling isolated or disconnected from the community that consists of the instructor and the learners. Lack of interaction among students and the instructor may negatively impact their learnings and cause adverse emotions like anxiety, sadness, and depression. Apart from the feeling of loneliness, sometimes students may come up with different issues or questions as they study the course, which can stop them from confidently progressing or make them feel discouraged if we leave them alone. To promote interaction and to overcome the limitations of geographic distance in online education, we propose a customized design, Tele-instruction, with useful features supplement to the traditional online learning systems to enable peers and the instructor of the course to interact at their conveniences once needed. The designed system can help students address their questions through the answers already provided to other students or ask for the instructor's point of view by two-way communication, similar to face-to-face forms of educational experiences. We believe our approach can assist in filling the gaps when online learning falls behind the traditional classroom-based learning systems.

## 1. INTRODUCTION

Nowadays, many universities, institutes, and companies offer online courses, training, certificates, or even degrees. Moreover, many people tend to educate online due to its advantages; they can educate from their home or their offices at their convenience time. Despite the fact that this temporal or geographical flexibility might help the learners manage their life in a better way, often, no face-to-face interaction occurs between students and the instructor.

Traditional learning environments let instructors use classroom interaction to assess student performance and engagement to utilize or adjust teaching strategies for improvements and learning satisfaction, whereas online learning platform which makes it difficult for the instructors to perform likewise due to lack of feedback from the students. Apart from that, many online learners may feel isolated or disconnected from the community, which can cause some adverse emotions like anxiety, sadness, and depression. In addition to the feeling of loneliness, from time to time some students may seek some clarifications on a concept, some may need a misunderstanding to be explained, and some others look for instructor's confirmation or influence. This could stop them from positively proceeding or make them feel discouraged if we leave them alone.

In order to promote effective learning and satisfaction in online education, there have been many studies, and as a result of those studies, many educational researchers have come up with different kinds of interaction as important and effective factors [6, 7, 19, 18, 25]. In most of the studies, however, three types of interaction, initially specified by Moore [20], i.e., learner to instructor (learner-instructor) interaction, learner to learner (learner-learner) interaction (known as peer interaction), and the interaction of the learner with content (learner-content), are considered as the main factors impacting student satisfaction and effective learning. While in some studies learner-instructor interaction plays the most vital role in student satisfaction, many others believe that learner-learner interaction might be a better predictor influencing learning outcomes. Some others emphasis on learner-content interaction instead [1, 5, 26, 14, 23, 13, 8, 17].

Our belief is that each type of interaction is essential and useful on its place so even if we, as researchers, spend our focus on one specific type of interaction or two at the moment, our ultimate goal is to integrate all features and have a customized design that helps in improvement of all three types of interaction. In this study, we propose a design, called Tele-instruction, that supports online learning with features that assist in filling the gaps when online learning falls behind the traditional learning systems. In particular, we would like to (i) promote effective interaction in online learning, and (ii) provide some features to help instructors and students benefit from feedback they obtain from each other.

We believe that a full-grown online instruction increases the capabilities of an online learning system by supplementing the possibility of face-to-face instant virtual interaction among the learning community, on demand. At the same time, all users enjoy pace and flexibilities promised by a typical online learning system. In Section 1.1, we review our related work. We then continue our discussions by reviewing the functionality of Tele-instruction and explaining its core modules, in Section 2. We conclude the work in Section 3, where we explain our future work.

## 1.1 Related work

Despite the rising popularity of online learning, their users mostly find it difficult to communicate with each other if questions or concerns arise, as communication is often very restricted, virtual, and offline. Articles [4, 16, 22] are some examples of the many studies confirming that the absence of face-to-face interaction between students and instructors negatively affects student performance. As also outlined in [9, 11, 12], there exists an essential need for design and development of online learning education technologies which promote effective interactivity and alleviate student loneliness feeling.

Considering online learning systems, one can distinguish between *synchronous* and *asynchronous* ways of interacting [3]. In asynchronous settings, learners can passively communicate with peers and possibly the instructor. Today, asynchronous online learning mostly involves email and message-passing services, or passive communications through forums and discussion boards. While this approach gives the students the opportunity of independency, it mostly works offline, and if the student is not able to talk to the point, back and forth message-passing would be needed until perception occurs. The perception process, however, could be very time consuming for both sides and, most of the time, the instructor receives several similar questions from different students taking the same course. In addition to long response time, utilizing forums and discussion boards may cause too much interaction among former students and lead to student overload, dissatisfaction, and backwardness [2, 18]. Last but not least, while by using asynchronous systems students can complete the course at their own pace, with the advantage of freedom from restrictions of space and time, students can still suffer from feeling isolated through the learning journey. On the other hand, using *synchronous* communication, students can interact with the instructor (and perhaps fellow students) instantly. Some designed systems offer "live chat" for almost real-time communications; Article [9], for instance, develops a tool for instant messaging and presence awareness that lets a student see if the instructor or another peer is online to chat. While this design allows students to interact with the instructor by message passing and informal chat once possible, students may have to stay online and check the status of others until it changes to "online", from either "away" or "offline" status. Although the designed tool can provide the feel of the community for peers and lets students communicate with the instructor informally, students might get busy by over-communication with peers typically happening in chat rooms, causing unsatisfactory, frustration, and other adverse learning outcomes.

Live chats ideally require 24/7 availability of someone familiar with the course concepts, like a teaching assistant or the instructor. However, since this full-time availability is not feasible or is costly, some studies implement the live chat using a robot, i.e., a chatbot. While this approach may provide faster feedback to the students, this communication does not help students stay away from feeling isolated and alone. Chatbots are not helpful in situations where students look for instructor's confirmation, clarification, or influence.

We propose a customized instructional design, called Tele-instruction, that supports interaction to the goal of effective learning in the online environment. Our system uses videoconferencing in a very flexible way, which provides real-time face-to-face interaction simulating the classroom environments. Simultaneously, students can enjoy the space and time-independency gifted by online education. Our design allows instructors to have adequate but not too much interaction with students which let instructors manage their time as they want, and helps students avoid feeling a lack of interaction. Instructors can use these face-to-face communications to earn feedback and adjust the course material, design assignments, and plan for appropriate course work and other activities similar to what they do in classroom learning systems. The designed system behaves as also emphasized in [21] and noted by Saba in [24]: "the success of distance education, to a great degree, will depend on the ability of educational institutions to personalize the teaching and learning process. Students should be given the chance to assess their comfort with the level of structure while learning at a distance and decide to what extent they need direct contact with the instructor (p.1)."

While some studies may share some similarities with our proposed system, like video conferencing, to the best of our knowledge, our system facilitates learning by taking advantage of some ideas that are not gathered as an integrated design in other existing related work.

## 2. TELE-INSTRUCTION

Tele-instruction's main idea of design is based on *appointments, videoconferencing, and learning management*. In the following sections, we explain the *core* modules of Tele-instruction:

## 2.1 Appointments

The instructor of the course specifies his/her available time slots (as short as a five-minute slot) for the upcoming week in a designed calendar, which can be shared with the students of the course. Students can access the calendar to book an appointment with the instructor. Meeting partners receive an automatic invitation to the online meeting as well as meeting reminders. Appointments can be merged if students have similar issues or concerns, which can provide the possibility of peer interaction and future peer discussions too. This merging could be offered to the instructor by the system or the instructor decides on that himself/herself.

Students who see the need for talking to the instructor have the option of choosing the best time that works for them among the times that the instructor is available. That helps both partners to go by their best comfort time and plan for the rest of their time at their own pace.

With these face-to-face communications, students receive feedback from the instructor. That not only helps them in a better understanding of the course concepts, it also gives them the feeling of inclusion, being heard and seen, and therefore satisfaction. Students can discuss their concern with the instructor and can resolve their misunderstanding of the concepts promptly. As also noted in [15], Northrup et al. are one of the many researchers who found instructor's feedback to the student as an important factor for student satisfaction in [18]. Bray et al., [6], also indicate that students who can easily communicate with their instructors perform better regarding learning compared to the ones who find it difficult to perform student-instructor interaction. As mentioned by Kuo in [15], "Learners in online environments report more course satisfaction when the support from their instructors matches their expectations of communicating with their instructors. Maintaining frequency of contact, having a regular presence in class discussion spaces, and making expectations clear to learners are three practices suggested for instructors to adopt in enhancing learner-instructor interaction during online learning". Similar findings are provided in [10]. Tele-instruction's Appointment module can provide an online learning system with the possibility of learner-instructor and effective learner-learner interaction.

The instructor has the chance to receive some feedback from the students, too, once they meet. That can help him behave similarly to what he does in classroom learning environments; he can use this feedback to adjust the course material and assessments accordingly. For example, if the instructor finds out that a concept is not understood enough by some students, he can set up some extra slides or examples explaining the same concept, or he can suggest some extra-reading material for those who are interested. See Section 2.6 for more on this matter.

## 2.2 Video conferencing and File Sharing

The instructor and students meet using the *Videoconferencing* feature of Tele-instruction. Students discuss the issue/concern with the instructor and get response correspondingly. The instructor may provide extra resources, like course slides, documents, or references to the student. Sometimes a student is needed to pass a training (attending a small lecture) in addition to the course material already provided. It could be, for example, to recall and master a background material prerequisite to the course. In this situation, the student has to accomplish the task, and the system allows the instructor to track students' progress.

## 2.3 Meeting Content Recording

The system provides the instructor with the option of recording the meeting contents. The recorded video/audio can be used by/recommended to the students for future needs as a lecture. Additionally, the recorded contents can be converted to text contents using a speech to text service to be used to address upcoming, similar questions. Stored contents will be accessible to the students with the same or similar questions via the *Learning Management System*.

## 2.4 Learning Management System

The learning management system works with the meeting contents. The idea is to manage the content and use it in the best way to answer students' questions. Once a student asks a question from the learning system, the system first looks for the best match through the available course material and guides the student by recommending the related material or answers correspondingly. If the answer is not available, the student will be guided to book an appointment to talk to the instructor of the course and get some feedback. The learning system gets trained as new contents come in.

## 2.5 Tracking progress and evaluations

Once the instructor or the learning management system assigns some learning material to be read by the student, the student's effort and progress can be tracked (instructor's choice) by the system to make sure the student has reviewed the contents. The student's knowledge can be evaluated afterward to make sure the issue has been resolved. The evaluations are helpful for the instructor to decide whether to change strategies or not.

## 2.6 Course design and instruction adjustments

Two-way communication between students and the instructor in in-person classes tells a lot to the instructor of the course. If the instructor receives so many questions regarding a concept while lecturing, he perceives that the concept is difficult for the students and some strategies are to be used to help students master that concept in a better way; like explaining the concept in more detail, solving more practice questions, preparing more assignments, or adding some extra lecture notes. Tele-instruction provides a similar opportunity for the instructor to adjust the lectures' speed and decide on teaching strategy while e-learning.

## 2.7 Trusted Technical Agents (TTAs)

Instructors may only need to assign a five-minute slot of their time to a student's meeting. This helps the student a lot, not only by educating the student but also mentally. Students feel like they are not left alone through the learning process, and they have the chance of being heard once needed. However, we need to make sure this feature does not ask a lot from the instructor of an online course. While we may need to arrange more appointments the first time we offer a course, eventually, the students will have most information they need access through the learning management system. This is because the meeting contents and instructor responses get recorded for future use. However, we know that many students are looking for interaction that helps them obtain the feeling of inclusion or instructor's confirmation or hints on what they do. Many of which indeed are their mental needs. addressing all such might not be feasible for the instructor of the course. So, for load-balancing purposes, the system makes it possible for the course instructor to introduce some trusted third parties, called *Trusted Technical Agents (TTAs)*, to help with the appointments, student guidance, and problem-solving once needed. TTAs could be course teaching assistants or students who are doing/have done well in the same course. Even if a student is doing well in the course and has already submitted an assignment, he would be a good TTA for helping other students regarding that assignment. Students can rate the agents through the system, and instructors can evaluate agents for the best performance.

Agents are pretty helpful once we have a large online class, and we want to give our students the possibility of having two-way communications simulating the same feeling they have in the traditional classroom environment.

## 3. CONCLUSIONS AND FUTURE WORK

We presented Tele-instruction, a customized design with the ultimate goal of improving three main types of interaction for learning satisfaction in online learning environments. Tele-instruction's appointments and videoconferencing procedures directly impact learner-instructor and learner-learner interaction. While as shown by many studies learner-learner and learner-instructor interaction can implicitly enhance learner interaction with content, by the use of Meeting Content module, which works on providing the contents of the meetings available to the students, we specifically aim at helping with enhancing effective student-content interaction. The quality of student-content interaction can be increased by assuring that the contents extracted from the meeting contents concentrate on comprehension, help students obtain a higher level of knowledge than what the course material provide, and finally work on improvement of the learner's thinking skills.

Using Tele-instruction, the student comes under the influence of a professional instructor, and as a result of that, the interaction of the student with the content will result in a better understanding of the course concepts, and therefore satisfaction. Learner-learner interaction can be supervised by an instructor during group-based appointments providing a higher level of peer interaction. This collaborative learning can continue after students get to know each other in the meetings, which results in more discussions and engagements for problem-solving conversations.

As a part of our future work, we develop a tool to incorporate Tele-instruction's modules into online learning systems. A tool with a well-designed dashboard can facilitate the instructor as well as students. Furthermore, we want to work on implementing automatic question classification features of the learning management system to help with better addressing student's questions using available material or deciding on leading the students to the calendar and book an appointment once needed. It would be interesting to elaborate on utilizing machine learning techniques to improve the system's performance for content comprehension.